\documentclass[aps,prl,twocolumn,superscriptaddress,showpacs,floatfix,nobibnotes]{revtex4}

\usepackage{epsfig}
\usepackage{epstopdf}

\usepackage{graphicx}
\usepackage{longtable}
\usepackage{CJK}
\usepackage{color}

\usepackage{mathptmx, courier, pifont}
\usepackage[scaled=0.92]{helvet}
\usepackage[T1]{fontenc}
\usepackage{textcomp}

\begin{document}

\begin{CJK*}{GB}{}

\title{Euclidean Dynamical Symmetry in Nuclear Shape Phase Transitions}

\author{Yu Zhang }
\affiliation{Department of Physics, Liaoning Normal University,
Dalian 116029, China}\affiliation{Department of Physics and State
Key Laboratory of Nuclear Physics and Technology, Peking University,
Beijing 100871, China}\affiliation{Department of Physics and
Astronomy, Louisiana State University, Baton Rouge, LA 70803-4001,
USA}

\author{Yu-{X}in Liu }
\affiliation{Department of Physics and State Key Laboratory of
Nuclear Physics and Technology, Peking University, Beijing 100871,
China} \affiliation{Center for High Energy Physics, Peking
University, Beijing 100871, China}

\author{Feng Pan}
\affiliation{Department of Physics, Liaoning Normal University,
Dalian 116029, China}\affiliation{Department of Physics and
Astronomy, Louisiana State University, Baton Rouge, LA 70803-4001,
USA}

\author{Yang Sun }
\affiliation{Department of Physics, Shanghai Jiao Tong University,
Shanghai 200240, China} \affiliation{Department of Physics and
Astronomy, University of Tennessee, Knoxville, Tennessee 37996,
USA}

\author{J. P. Draayer}
\affiliation{Department of Physics and Astronomy, Louisiana State
University, Baton Rouge, LA 70803-4001, USA}

\date{\today}

\begin{abstract}
The Euclidean dynamical symmetry hidden in the critical region
of nuclear shape phase transitions is revealed by a novel algebraic
F(5) description. With a nonlinear projection, it is shown that the
dynamics in the critical region of the spherical--axial deformed and
the spherical--$\gamma$ soft shape phase transitions can indeed be
manifested by this description, which thus provides a unified
symmetry--based interpretation of the critical phenomena in the
region.
\end{abstract}
\pacs{21.60.Fw, 21.60.Ev, 21.10.Re, 64.70.Tg}

\maketitle

\end{CJK*}

%\newpage

Dynamical symmetries (DSs) play an important role in elucidating the
quintessential nature of quantum many-body dynamical structures,
especially their evolution under changing conditions. Typical
examples of DS are those associated with the interacting boson model
(IBM)~\cite{IachelloBook87} for nuclear structure and the vibron
model (VM)~\cite{IachelloBook95} for molecules and atomic
clusters~\cite{HJP2006}, where various DSs provide considerable
insight into the nature of shape phases and shape phase transitions
(SPTs)~\cite{Cejnar2010}.

The IBM possesses an overall U(6) symmetry with three {DSs}
corresponding to three special nuclear shapes or collective modes;
namely, a spherical vibrator [U(5)], an axially deformed rotor
[SU(3)], and a $\gamma$-soft rotor
[O(6)]~\cite{ICSM-Initial,ICSM-Rot}. In nuclei, the typical shape
phase diagram can be characterized by the so-called Casten
triangle~\cite{Casten2007} in the IBM parameter space with the three
DSs placed at the vertices of the triangle as shown in
Fig.~\ref{F2}. Experimental observations show not only that {these
three DSs indeed exist in nuclei, but also the SPTs occur with two
good} examples~\cite{IachelloBook87} being the first-order SPT from
U(5) to SU(3) and the second-order SPT from U(5) to O(6).
Additionally, quasidynamical symmetries have been found to occur
along the legs of the Casten triangle~\cite{Rowe2004} and even
inside the triangle~\cite{Bonatsos2010}. It has been also shown that
{partial dynamical symmetries may occur at the critical point of a
SPT~\cite{Leviatan2007}. On the other hand, within the
Bohr-Mottelson Model (BMM)~\cite{Bohr1998}, the
E(5)~\cite{Iachello2000,Caprio2007} and X(5)~\cite{Iachello2001}
critical point symmetries (CPSs) were developed to approximately but
analytically describe the states at the critical point of the
spherical to $\gamma$-soft SPT and those of the spherical to axially
deformed SPT, respectively. Accordingly, the structural paradigms in
the triangle shown in Fig.~\ref{F2} can also be labelled with the
BMM language of vibrator, (axial) rotor, $\gamma$-soft (rotor), E(5)
and X(5), which are the solutions to the Bohr Hamiltonian. However,
the distinction between the IBM and the BMM should be borne in mind.
Then the algebraic collective model was developed to provide a
computationally tractable version of the BMM~\cite{ACM1}. However,
the DS structure of the CPSs is still lacking. In this work, we will
make clear the dynamical structure of the CPSs, and establish the
approach to describe the states in the transitional region
connecting the two critical point symmetries as shown in
Fig.~\ref{F2} in a unified way.

The E(5) CPS was initiated with the solution of the five-dimensional
square well potential in the BMM, and the corresponding Hamiltonian
is invariant under both translations and rotations in
five-dimensional space if confined in the well. It holds then the
five-dimensional Euclidean symmetry, the Eu(5)
symmetry~\cite{Caprio2007}. By implementing $d$-boson creation and
annihilation operators for the five-dimensional system with
\begin{equation}
\tilde{d}_{u} =\frac{1}{\sqrt{2}}\left[q_u+i\tilde{p}_u\right],
~~~~ d^{\dag}_{u} =\frac{1}{\sqrt{2}}\left[q_u - i
\tilde{p}_{u}\right]\, ,
\end{equation}
where $q_u$ and  $\tilde{p}_u$  are the coordinates and the
associated momenta in spherical tensor
form~\cite{Marshalek2006,Klein1982} with $\tilde{A}_u=(-)^uA_{-u}$,
and using the definition of Casimir operator of the Eu(5) group,
$C_{2,\textrm{\footnotesize{Eu(5)}}}=\tilde{p}^2$ (see, for example,
Ref.~\cite{Caprio2007}), one can give the algebraic Hamiltonian with
the Eu(5) DS,
\begin{equation}\label{HF(5)}\hat{H}_{\rm F(5)}=\alpha
[\hat{n}_{d} + \frac{5}{2} - \frac{1}{2}\left(\hat{P}^{\dag}_{d} +
\hat{P}_{d} \right)]\, ,
\end{equation}
where $\alpha$ is a scale factor, $\hat{n}_{d}=\sum_{u} d^{\dag}
_{u} d_{u}$ and $\hat{P}_{d} = \sum_{u} (-)^{u} d_{u} d_{-u}$. The
Hamiltonian (\ref{HF(5)}) can be diagonalized~\cite{PD} under the
$\mathrm{U(5)}\supset \mathrm{O(5)}\supset \mathrm{O(3)}$ basis $\{|
n_d \, \tau \; \Delta \, L \rangle\}$ with $0\leq n_{d}<\infty$. It
should be noted that this scheme does not lie in the framework of
the IBM due to the non-compactness of the Eu(5) group, but can be
translated directly from the geometric description of the E(5)
CPS~\cite{Iachello2000} because the Hamiltonian of the latter may be
written as $\hat{H}_{\rm
BMM}^{}=\frac{1}{2B}\tilde{p}^2=\frac{1}{2B}C_{2,\textrm{\footnotesize{Eu(5)}}}$,
which, however, should be confined within an infinite square
well~\cite{Caprio2007}. Besides, it is not easy to include the
boundary condition of the square well directly in the algebraic
realization when diagonalizing the Hamiltonian~(\ref{HF(5)}). Owing
to the fact that the boson number $N$ is fixed in the IBM, if the
$d$-bosons constructed in (1) are regarded to be equivalent to those
in the IBM, practical calculation with the algebraic Eu(5)
Hamiltonian~(\ref{HF(5)}) can be realized by diagonalizing the
corresponding IBM analogue within the U(6) subspace for fixed boson
number $N$. We refer it then to the F(5) scheme. One can verify
numerically that ratios of the eigen-energies and the eigenstates of
an infinite well problem can indeed be produced approximately by
diagonalizing the Hamiltonian (\ref{HF(5)}) within a finite boson
subspace. The larger the boson number $N$, the better the
approximation. Thus, the link between the geometric and the
dynamical F(5) algebraic description of the CPS in the critical
region of the SPT is established.

\begin{figure}
\begin{center}
\includegraphics[scale=0.35]{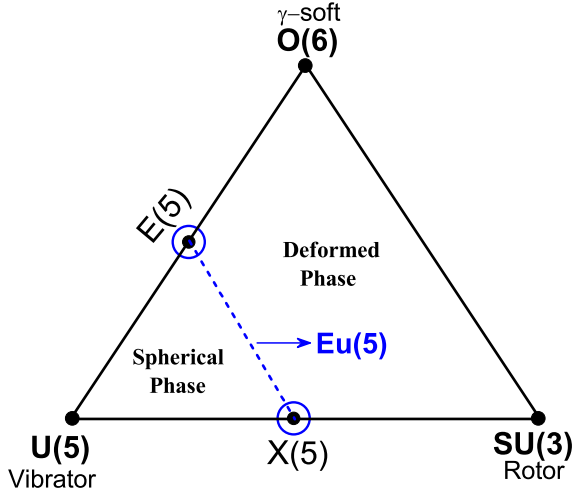}
\caption{(Color online) Nuclear shape phase diagram characterized by
the symmetry triangle. Note that there are two systems for labeling
this paradigm; the geometric language and the IBM (see the text).
\label{F2}}
\end{center}
\end{figure}

\begin{table*}[htb]
\caption{Typical energy and $B(E2)$ ratios in the F(5) scheme with
different $\chi$ at $N=1000$ together with those in the related
models.} \label{T1}
\begin{tabular}{cccccccccccc}\hline\hline
&~~E(5)~~&\multicolumn{6}{c} {F(5) at $N=1000$}
&~~X(5)~~&~~U(5)~~&~~O(6)~~&~~SU(3)~~ \\
\cline{3-8} & &$\chi=0.0$&$\chi=-0.4$&$\chi=-0.8$&$\chi=-1.0$ & $\chi=-1.1$ &
$\chi=-1.32$&\\ \hline
$E_{4_1}/E_{2_1}$&2.20&2.19&2.33&2.51&2.63&2.71&2.89&2.91&2.00&2.50&3.33\\
$E_{6_1}/E_{0_2}$&1.18&1.19&1.12&1.05&1.01&1.00&0.96&0.96&1.50&1.00&$\frac{21}{8N-4}$\\
$E_{0_2}/E_{2_1}$&3.03&3.02&3.53&4.22&4.67&4.93&5.61&5.67&2.00&4.50&$\frac{4(2N-1)}{3}$\\
$\frac{B(E2;~4_1\rightarrow 2_1)}{B(E2;~2_1\rightarrow 0_1)}$ ~
&1.68&1.67&1.65&1.63&1.62&1.61&1.60&1.58&$\frac{2(N-1)}{N}$
&$\frac{10(N^2+4N-5)}{7(N^2+4N)}$&$\frac{10(4N^2+6N-10)}{7(4N^2+6N)}$\\
$\frac{B(E2;~0_2\rightarrow 2_1)}{B(E2;~2_1\rightarrow 0_1)}$ ~
&0.86&0.86&0.79&0.72&0.68&0.66&0.62&0.63&$\frac{2(N-1)}{N}$&0.00&0.00\\
\hline\hline
\end{tabular}
\end{table*}

The E(5) and X(5) models are both restricted {to an infinite
square well potential in $\beta$, the only difference between the
two models is how the $\gamma$ degree of freedom is
handled~\cite{Iachello2000,Iachello2001}.} If only $n_\gamma=0$
states in the X(5) model~\cite{Iachello2001} are considered, which
{corresponds to the yrast and yrare states, the $\beta$ dependence
in the E(5) and X(5) models} can be expressed uniformly by the
Bessel equation:
\begin{equation} \label{Bessel-Eq}
\psi^{\prime\prime}(z)+\frac{\psi^\prime(z)}{z}+\left(1-\frac{v^2}{z^2}\right)\psi(z)=0
\, ,
\end{equation}
where $\psi(z)\sim z^{-3/2}J_v(z)$ {with $J_{v}(z)$ being a Bessel
function of order $v$, in which $z$ is proportional} to the $\beta$
variable. For the E(5) model, $v=\tau+3/2$ with $\tau$ being the
seniority number of the O(5) group, while for the X(5) model, $v =
\left[\frac{L(L+1)}{3}+ \frac{9}{4}\right]^{1/2}$ with $L$ being the
{angular momentum quantum number.} Accordingly, we can establish a
{mapping} $v=f(L,\chi)$ with $f(L,0)=\tau$ since $L=2\tau$ for the
yrast states in this case and  $f(L,-\frac{\sqrt{7}}{2})
=\left[\frac{L(L+1)}{3}+ \frac{9}{4}\right]^{1/2}$.
Obviously, there are many different choices {for $f$, but since
they are homotopic to one and another, each one should then correspond to a way
to get those from} the E(5) critical point to the X(5) critical point.
For simplicity, we take the linear {mapping}
\begin{equation}  \label{v-general}
v=\left( 1 \! + \! \frac{2}{\sqrt{7}}\chi \right)
\frac{L}{2}-\frac{2\chi}{\sqrt{7}} \left[\frac{-3 \! + \! \sqrt{9 \!
+ \! 4L(L\! + \! 1)/3}}{2} \right] + \frac{3}{2} \,
\end{equation}
with $\chi\in[0,-\frac{\sqrt{7}}{2}]$. For a given $\chi$, we
define} a projection $\hat{P}_{\tau^\prime,\tau}^\chi$ that projects
the quantum number $\tau$ to be equivalent to
$\tau^{\prime}=\nu-3/2$ according to (\ref{v-general}). Obviously,
the projection is nonlinear because of the nonlinear dependence of
$\nu$ on the quantum number $L$ ($\tau$) shown in (\ref{v-general}).
We found that, after the projection, the Hamiltonian given in (5)
can be rewritten in terms of functionals of the U(5) operators with
\begin{eqnarray}\label{PHF(5)}
\!\!& \!\! &\!\!\!\! \hat{H}^{\prime}_{\rm
F(5)}=(\hat{P}_{\tau^\prime,\tau}^\chi)^\dagger\hat{H}_{\rm F(5)}
\hat{P}_{\tau^\prime,\tau}^\chi= \nonumber\\
&& \; \; A+\frac{2\chi}{\sqrt{7}}\sqrt{B}-\frac{\chi}{\sqrt{7}}\sqrt{\frac{16}{3}B
-\frac{40}{3}\sqrt{B}+17}+\frac{5}{2}\nonumber\\
&& \; \;
-\frac{A+(1+\frac{4\chi}{\sqrt{7}})\sqrt{B}-\frac{2\chi}{\sqrt{7}}\sqrt{\frac{16}{3}B
-\frac{40}{3}\sqrt{B}+
17}+\frac{7}{2}}{2(A+\sqrt{B}+\frac{7}{2})}~C^\dag\nonumber\\
&& \;\;
-~C\frac{A+(1+\frac{4\chi}{\sqrt{7}})\sqrt{B}-\frac{2\chi}{\sqrt{7}}\sqrt{\frac{16}{3}B
-\frac{40}{3}\sqrt{B}+17}+\frac{7}{2}}{2(A+\sqrt{B}+\frac{7}{2})},
\;\;\;
\end{eqnarray}
where $A=\hat{n}_d$, $B=\hat{n}_d(\hat{n}_d+3)-2P_d^\dag
P_d+\frac{9}{4}$, $C=P_d$, and the scale factor in (\ref{HF(5)}) has
been set as $\alpha=1$. The expression~(\ref{PHF(5)}) is the
Hamiltonian for $\chi\in[0,-\frac{\sqrt{7}}{2}]$, which is well
defined when being diagonalized under the
$\mathrm{U(6)}\supset\mathrm{U(5)}\supset \mathrm{O(5)}\supset
\mathrm{O(3)}$ basis, and regains the Hamiltonian~(\ref{HF(5)}) as
taking $\chi=0$. The quadrupole operator in this case may be {taken
simply as} $T_u=e(d^\dag+\tilde{d})_u^{(2)}$ with $e$ being an
effective charge. As a result, a symmetry-based realization of the
dynamical structural evolution between the E(5) and the X(5) CPSs is
provided in the F(5) scheme.

Several typical {energy and $B(E2)$ ratios} in the related models
are listed in Table~\ref{T1}. {The results show clearly} that the
F(5) scheme with $\chi = 0$ and $\chi = -1.32$ in the large $N$
limit reproduces nicely the results of the E(5) and X(5) models.
Furthermore, the calculated quantities increase or decrease
monotonously {as $\chi$ changes from the X(5) limit with
$\chi\approx-1.32$ to the E(5) limit with $\chi=0$, which all
fall} between those of the spherical vibrator [U(5)] and the
deformed rotor [O(6), SU(3), or O(6) and SU(3) mixed for some
cases]. The results indicate that the Eu(5) DS can definitely be considered
as the critical DS of the spherical to deformed SPT region as
shown in Fig.~\ref{F2}.

It is remarkable that the bandhead energies of excited $0^+$ states
for any given $N$ in the F(5) scheme are universally independent of
$\chi$ when normalized to $E_{0_2}$. For example,
$E_{0_3}/E_{0_2}=2.57$ for $N=10$ and $E_{0_3}/E_{0_2}=2.50$ for
$N=1000$, which in the large $N$ limit coincides with the rule of
$E_{0_{n}}=A(n-1)(n+2)$~\cite{Bonatsos2008II}, where $A$ is a
$\chi$-dependent parameter. The analysis in
Ref.~\cite{Bonatsos2008II} shows that the same law also occurs to
the excited $0^+$ states around the critical point of the
U(5)--SU(3) SPT in the large $N$ limit. Similarly, energies of the
excited $14^+$ states in the F(5) scheme are also independent of
$\chi$ for any given $N$. This can be easily explained based on
(\ref{v-general}), in which the values of $v$ for $L=0$ and $L=14$
are independent of $\chi$ and given by $v=\frac{3}{2}$ and
$v=\frac{17}{2}$, respectively. As a result, the ratio
$E_{14_1}/E_{0_2}$ can be taken as a signal of the Eu(5) DS
occurring in even-even nuclei. Furthermore, as shown in
Table~\ref{T1}, energies of the $6^{+}_{1}$ and $0^{+}_{2}$ states
in the F(5) scheme with $-1.32\leq\chi\leq-0.8$ are approximately
degenerate in the large $N$ limit. Detailed calculations indicate
that the approximate degenerate situations also occur among other
states, e.g., ($10^{+}_{1}$, $0^{+}_{3}$), ($14^{+}_{1}$,
$0^{+}_{4}$), and so on, but the degeneracies are gradually removed
with {increasing excitation energies.} As discussed in
Ref.~\cite{Bonatsos2008}, the degeneracies of ($6^{+}_{1}$,
$0^{+}_{2}$), etc, are the signature of the underlying symmetry
within the critical region in the large $N$ limit. These numerical
results demonstrate further that the underlying symmetry can be
attributed to the Eu(5) DS at least for low-lying states. Moreover,
the experimental data of some typical quantities for the
transitional nuclei~\cite{Cejnar2010,Casten2007} previously
identified as the candidates of either the E(5)~\cite{Iachello2000}
or X(5) CPS~\cite{Iachello2001}, together with those calculated from
Eq.~(\ref{PHF(5)}), are shown in Table~\ref{T2}. One can observe
from Table~{\ref{T2}} that the experimental data are well fitted by
the F(5) scheme except for the inter-band $B(E2)$ ratio shown in the
last row. A possible improvements in the theoretical prediction
about the inter-band $B(E2)$ ratio may be made by adding additional
terms such as $(d^\dag\tilde{d})_u^{(2)}$ in the quadrupole operator
based on the analysis in \cite{Arias2001}. More specifically, the
approximate degeneracy of the $6^{+}_{1}$--$0^{+}_{2}$ levels
emerges clearly both in experiment and in the F(5) scheme for the
cases with large $|\chi|$ value, and the constant value of
$E_{14_1}/E_{0_\xi}$ predicted by the theory is also confirmed {by}
experiment, which indicates that these critical nuclei may be
possible candidates {for} the Eu(5) symmetry.
\begin{table}[htb]
{\scriptsize\caption{Some typical energy and $B(E2)$ ratios for
$^{102}$Pd~\cite{Frenne9809,Zamfir2002},
$^{128}$Xe~\cite{Kanbe2001,Coquard2009},
$^{146,148}$Ce~\cite{Peker1997,Bhat2000}, and
$^{150}$Nd~\cite{Basu2013}, together with those calculated from
Eq.~(\ref{PHF(5)}) with different $\chi$ at $N=1000$, where ``$-$''
denotes the quantities undetermined {in} experiment. (N.B., the
$0_{\xi}$ state represents the band head state of the $\xi=2$ family
as that in the E(5) CPS description.)} \label{T2}
\begin{tabular}{cccccc}\hline
Ratio &&&($\chi$,~nucleus)&&\\
\cline{2-6}&($-0.2$,$^{102}$Pd)&($-0.4$,$^{128}$Xe)&($-0.9$,$^{146}$Ce)&($-1.2$,$^{148}$Ce)&($-1.3$,$^{150}$Nd)\\\hline
$E_{4_1}/E_{2_1}$&(2.26,~~2.29)&(2.33,~~2.33)&(2.57,~~2.58)&(2.78,~~2.86)&(2.89,~~2.93)\\
$E_{6_1}/E_{2_1}$&(3.75,~~3.79)&(3.94,~~3.92)&(4.58,~~4.53)&(5.13,~~5.30)&(5.40,~~5.53)\\
$E_{8_1}/E_{2_1}$&(5.46,~~5.42)&(5.80,~~5.67)&(6.95,~~6.72)&(7.94,~~8.14)&(8.44,~~8.68)\\
$E_{6_1}/E_{0_{\xi}}$&(1.15,~~1.27)&(1.12,~~~0.97)&(1.03,~~1.12)&(0.98,~~1.09)&(0.96,~~1.07)\\
$E_{14_1}/E_{0_{\xi}}$&(3.64,~~3.70)&(3.64,~~~2.57)&(3.64,~~$-$~~)&(3.64,~~3.75)&(3.64,~~3.97)\\[3pt]
$\frac{B(E2;~4_1\rightarrow 2_1)}{B(E2;~2_1\rightarrow 0_1)}$
&(1.66,~~1.56)&(1.65,~~1.47)&(1.62,~~$-$~~)&(1.61,~~$-$~~)&(1.60,~~1.56)\\[4pt]
$\frac{B(E2;~0_{\xi}\rightarrow 2_1)}{B(E2;~2_1\rightarrow 0_1)}$
&(0.83,~~0.39)&(0.79,~~0.33)&(0.70,~~$-$~~)&(0.64,~~$-$~~)&(0.62,~~0.37)\\
\hline
\end{tabular}}
\end{table}

\begin{figure}[htb]
\begin{center}
\includegraphics[scale=0.32]{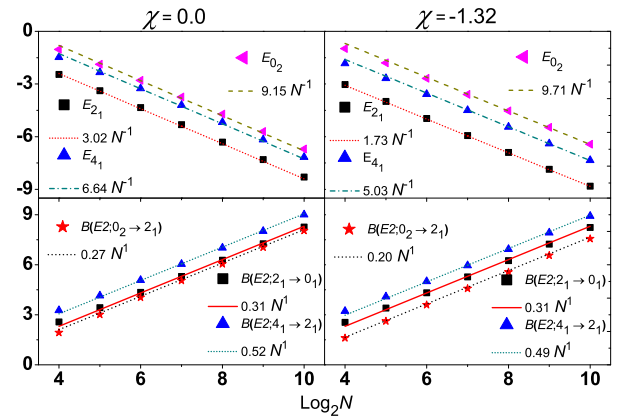}
\caption{Scaling behaviors of some typical energies and $E2$
transition rates with respect to $N$ (log$_2$ by log$_2$) for the
F(5) scheme with two $\chi$ cases.} \label{F3}
\end{center}
\end{figure}

We also {investigated} the scaling properties of some typical
quantities in the F(5) scheme for the case with $\chi=0$
corresponding to the E(5) and with $\chi=-1.32$ corresponding to the
X(5) critical point. The results are shown in Fig.~\ref{F3}. It is
{also evident from} Fig.~\ref{F3} that each excited level scales
with $N^{-1}$, and each $E2$ transition rate scales with $N^1$.
Along the analysis in Ref.~\cite{Rowe2004}, if a Hamiltonian
$H=-\bigtriangledown^2/(2M)+k\beta^{2n}$ with $k\propto M^{~t}$, its
spectrum should have a scale factor $M^{(t-n)/(n+1)}$. Therefore,
the spectrum of an infinite square well should have a scale factor
$M^{(t-n)/(n+1)}\mid_{n\rightarrow\infty}=M^{-1}$. The $N^{-1}$
power law of the spectrum in the F(5) scheme is indeed consistent
with the conclusion with $M\propto N$ as shown in
Ref.~\cite{Rowe2004}. It is apparent that ratio of two quantities
must be an $N$-independent constant if they obey the same power law.
As a result, the $N$-scaling law of the F(5) scheme shows that the
Eu(5) DS is well kept in finite $N$ cases, which in turn suggests
that the CPS associated with an infinite well is robust in finite
systems.

In summary, we {proposed an algebraic F(5) scheme to reveal the
hidden Eu(5) DS} in the critical region of the spherical--deformed
SPTs. It provides thus a new perspective to understand the nuclear
dynamics in the transitional region. {We have} shown that the Eu(5)
DS can be directly translated from the geometric description of the
CPS of the U(5)--O(6) transition. With the nonlinear projection, the
structural evolution from the CPS of the U(5)--O(6) to that of the
U(5)--SU(3) transition is realized. Our numerical analysis shows
that the experimental data are reproduced well in the scheme, which
indicate that the Eu(5) DS is dominant but hidden in the whole
critical region of the SPT.

\bigskip

\begin{acknowledgments}
The authors are thankful to Dr. J. N. Ginocchio for illuminating
discussions. Work supported by the National Natural Science
Foundation of China under Contract Nos. {11375005, 11005056,
11175078, 10935001, 11075052 and 11175004}, the National Key Basic
Research Program of China under Contract No. G2013CB834400, the
Doctoral Program Foundation of State Education Ministry of China
under Contract No. {20102136110002}, the U.S. {National Science
Foundation under Contract No. OCI-0904874,} the Southeastern
Universities Research Association, and the LSU--LNNU joint research
program under Contract No. 9961.
\end{acknowledgments}

%\newpage

%\section*{References}

\end{document}